# Observation of Thermal Deuteron-Deuteron Fusion in Ion Tracks

K. Czerski[1], R. Dubey[1], A. Kowalska[2], G. Das Haridas[1], S. Thulichery[1], M. Kaczmarski[1], N. Targosz-Sleczka[1], M. Valat[1]

[1] Institute of Physics, University of Szczecin, 70-453 Szczecin, Poland

[2] Physics Department, Maritime University of Szczecin, 70-500 Szczecin, Poland

**Abstract:**

*A direct observation of the deuteron-deuteron (DD) fusion reaction at thermal meV energies, although theoretically possible, has not succeeded up to now. The electron screening effect that reduces the repulsive Coulomb barrier between reacting nuclei in metallic environments by several hundreds of eV and is additionally increased by crystal lattice defects in the hosting material, leads to strongly enhanced cross sections which means that this effect might be studied in laboratories. Here we present results of the $^2H(d,p)^3H$ reaction measurements performed on deuterated metallic (Zr, Ti and Pd) targets down to the lowest deuteron energy in the center mass system of 625 eV, using an ultra-high vacuum accelerator system. The experimental thick target yield, decreasing over seven orders of magnitude for lowering beam energies, could be well described by the electron screening effect and recently found narrow threshold resonance in $^4He$. At the lowest beam energies, a constant reaction yield plateau could be observed for all studied materials. As indicated by significantly increased energies of emitted protons, this effect can be associated with the thermal DD fusion. A theoretical approach based on the thermal spike model explains the experimental observations by creation of ion tracks induced in the target by projectiles, and a high phonon density, which locally increases temperature above the melting point. Comparison of yield plateaus observed for the used target materials of different physical properties confirms the importance of the enhanced deuteron diffusion in ion tracks as well as electron screening and the threshold resonance in $^4He$. The experimental results also enable determination of the reaction rates and cross section at thermal energies, opening new perspectives for both astrophysical and commercial applications.*

## 1. Introduction

Nuclear reactions between charged particles at very low energies are determined by the penetration probability through the Coulomb barrier, which leads to dropping reaction cross section for lowering projectile energies. Its height can be, however, significantly reduced due to the inevitable presence of surrounding electrons. This so-called electron screening effect can be observed for single atoms, molecules and condensed-matter conditions and results in an exponential-like enhancement of reaction cross sections at decreasing energies when compared to the bare nuclei case [1]. A particularly strong effect could be observed for metallic environments in both direct [2,3,4] and inverse reaction kinematics [5], where quasi-free conduction electrons make the largest contribution. Therefore, these experiments were considered as model studies of electron screening in strongly coupled astrophysical plasma, where the thermal energy of the constituents is lower than the average Coulomb repulsion energy. Such plasma is found in white, red and brown dwarfs as well as in giant gas planets [6,7] representing an example of either superdense or relatively cold matter. In the latter case, deuteron-deuteron (DD) fusion reactions are of particular interest due to their relatively low Coulomb barrier of around 350 keV and expected important energy source, significantly increasing the core temperature of planets such as Jupiter and Saturn [8,9]. This was supported by the finding that the exponential enhancement of the reaction cross section experimentally observed for lowering deuteron energies, described by the electron screening energy, was much stronger than predicted by theoretical models [10,11]. Only in the last few years it could be shown that this effect results due to localization of electrons by crystal lattice defects and small oxygen impurities in the irradiated metallic targets [12,13]. Investigations performed on Zr target, which strongly binds hydrogen and produces a stable $ZrD_2$ stoichiometry, delivered screening energies ranging between 100 eV for a defect free target and up to 500 eV for the Zr target implanted by small oxygen doses [13]. Whereas the lowest screening energy agrees with the theoretical value [10], the higher values arise due to crystal lattice vacancies decorated by oxygen and hydrogen, which was confirmed using positron annihilation spectroscopy [12,14].

Surprisingly, recent accelerator studies of low-energy DD fusion have shown that its enhancement also results from a threshold 0+ resonance in $^4$He [15,16]. Its signature could be observed in the $^2$H(d,p)$^3$H reaction based on destructive interference with well-known broad resonances in $^4$He [17]. In agreement with theoretical predictions, this resonance was found to decay predominantly by positron-electron pair creation, which was recently confirmed by the observation of high energy bremsstrahlung and annihilation radiation [18,19]. A direct consequence of these investigations is that the fusion of two deuterons can take place with probabilities that enables its observation at thermal meV energies [16].

In this paper, we present new measurements of the $^2$H(d,p)$^3$H reaction performed for the first time at unprecedentedly low deuteron beam energies applying a deuterated Zr, Ti and Pd targets and the recently upgraded accelerator system with ultra-high vacuum at the University of Szczecin, Poland. It is expected that electron screening exponentially enhances the reaction cross sections for lowering projectile energies due to increase of the penetration factor which can be estimated for the s-wave partial wave as follows:

$$P_{scr}(E + U_e) = \sqrt{\frac{E_G}{E+U_e}} \exp\left(-\sqrt{\frac{E_G}{E+U_e}}\right), \quad (1)$$

where the screening energy $U_e$ has been added to the center mass energy $E$ of reacting nuclei. The Gamow energy $E_G = 2(Z_1 Z_2 \pi/137)^2 \mu$ is equal to 986 keV for the DD system ($\mu$ stays for the reduced mass). The reaction cross section can be then obtained using the expression for the penetration factor and the astrophysical S-factor $S(E)$:

$$\sigma_{scr}(E) = \frac{1}{\sqrt{E(E+U_e)}}\, S(E)\, \exp\left(-\sqrt{\frac{E_G}{E+U_e}}\right) = \frac{1}{\sqrt{E_G\, E}}\, S(E)\, P_{scr}(E+U_e) \qquad (2)$$

According to the formulas above, the cross section of the $^2$H(d,p)$^3$H reaction at $E$=1 keV (corresponding to 2 keV in Lab system) should be enhanced compared to the bare nuclear case ($U_e = 0$) by a factor 4.3 or 320 depending on the screening energy equal to 100 or 500 eV, respectively. In the previous experiment on Zr [15], $E$=3 keV could be reached, and the enhancement factor of about 2 could be observed. In this work, we will show that the experimentally measured reaction yield no longer decreases for the lowest deuteron energies, forming a characteristic yield plateau and leading to enhancement factors of many orders of magnitude. Moreover, the observation of emitted protons with energies significantly higher than those predicted by DD fusion kinematics indicates that they originate in a center-of-mass system that is nearly at rest.

This effect will be explained by the emission of protons from ion tracks induced by the deuteron beam, which, due to cascade collisions with target atoms, create cylindrical areas with high temperatures, exceeding even the melting point of the target material at the beginning of the ion track evolution. The high temperature in ion tracks leads to a strongly increased diffusion of deuterons implanted in the target and a higher number of DD collisions on meV energy scale. The ion track formation as a collision cascade can be described within the thermal spike model, which was developed in the 1970s to explain sputtering of a target surface by low energy ions [20-22]. Later, this model was also extended to describe interaction of swift heavy ions with matter, introducing separate temperatures for the electron gas and the lattice atoms, using electron-phonon coupling as the heat transfer mechanism [23,24]. The same model can also be applied to laser irradiation with short pulses (femtosecond/picosecond) causing crystal defects, usually supplemented with molecular dynamics studies to track the motion of atoms [25,26].

The new measurements of the $^2$H(d,p)$^3$H reaction have been conducted on various target materials characterized by distinct physical parameters, such as heat capacity, thermal conductivity, deuteron diffusion, and electron screening energies. This will enable not only to test the ion track model of the thermal proton emission, but also to formulate conditions for which the studied effects can be significantly enhanced.

(a) (b)

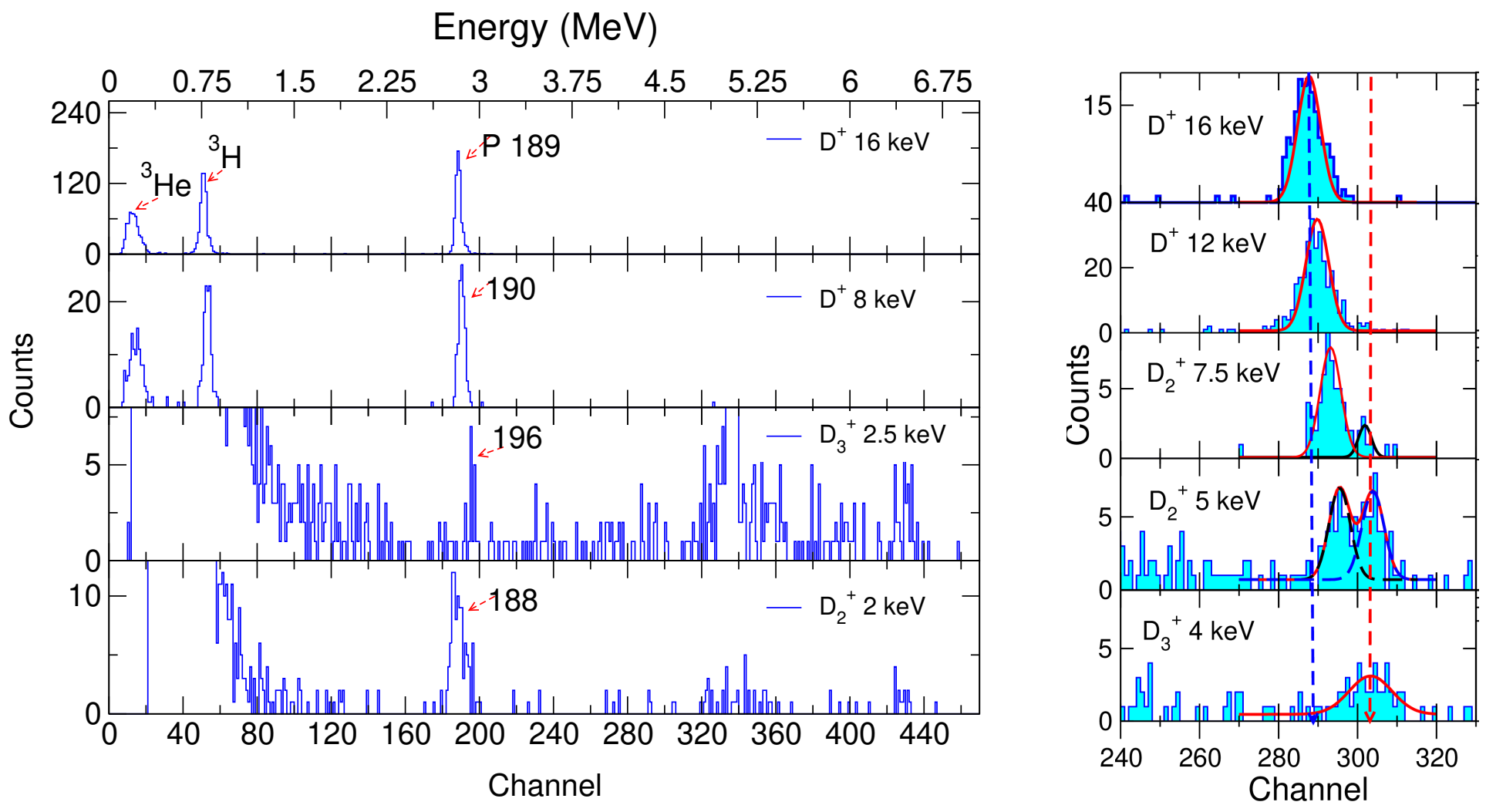

**Fig. 1:** Charged particle spectra of the $^2H(d,p)^3H$ reaction at different deuteron energies, applying atomic and molecular deuterium beams. **(a)** The spectrum measured on Zr. The numbers at the proton line show the channel number of the peak position. **(b)** The 3 MeV proton line observed in measurements performed on Pd. Due to a better energy resolution two spectral components could be observed. The low-energy component decreases with lowering projectile energies, and its peak position corresponds to the DD fusion kinematics. The high-energy component is characterized by a fixed maximum position and constant amplitude; it arises from proton emission in the near resting center mass system.

## 2. Results

Measurements at the lowest deuteron energies conducted for $ZrD_2$ and $TiD_2$ targets were possible due to utilizing of a new deceleration lens system, allowing us to operate the ECR ion source of the accelerator system at relatively high voltages with high beam currents. In the case of deuterated Pd target, the reaction yield plateau has been observed at higher deuteron energies, thus the deceleration system has not been necessary to use. Atomic, $D_2^+$ and $D_3^+$ molecular ion beams have been applied to determine the thick target yield in a broad energy range of deuterons between 1.25 keV and 25 keV (0.625 keV and 12.5 keV in CMS). A low noise PIPS Si detector placed backwards to the beam at 135° was employed to measure all charged particles emitted: protons, tritons, $^3He$ particles and electrons/positrons, produced by the DD fusion reactions. In front of the detector, an aluminum foil of different thicknesses was placed to prevent elastically scattered deuterons from entering the detector. The experimentally determined energy spectra are presented in Fig. 1.

As targets, metallic Zr, Pd and $Ti_{0.993}Pd_{0.007}$ plates of 1 mm thickness were used. They were implanted with deuterons of different beam energies to a saturation level, leading to a high stoichiometric (deuterium to metal atom) ratio of 2 for zirconium and titanium targets, which strongly bind deuterium leading to small diffusion coefficients. In the case of palladium, the stoichiometric ratio amounted only to 0.05 due to a weak hydrogen binding, correlated with a large deuterium diffusion (see Table 1 and 2).

### 2.1 Low-energy thick target yield

The experimentally measured thick target yield was calculated as the number of detected protons divided by the deuteron beam current (expressed in μA) and the measurement time (in thousands of seconds). On the other hand, it can be theoretically determined using the expression for the screened reaction cross section and the integral over the target depth up to the projectile range $R$:

$$Y_{scr}(E) = N_0 \int_0^R \sigma_{scr}(E)\, dx = N_0 \int_0^E \frac{\sigma_{scr}(E)}{|dE/dx|} dE \simeq \frac{2N_0\, S(E)}{C\sqrt{E_G}} \exp\left(-\sqrt{\frac{E_G}{E+U_e}}\right) \quad (3)$$

where the stopping power function in the energy region studied is governed by the electronic stopping $|dE/dx| = C\sqrt{E}$ . $N_0$ stays for the deuteron number density in the target that could be experimentally estimated at the highest beam energies (above 15 keV) for which the contribution of the electron screening effect can be neglected.

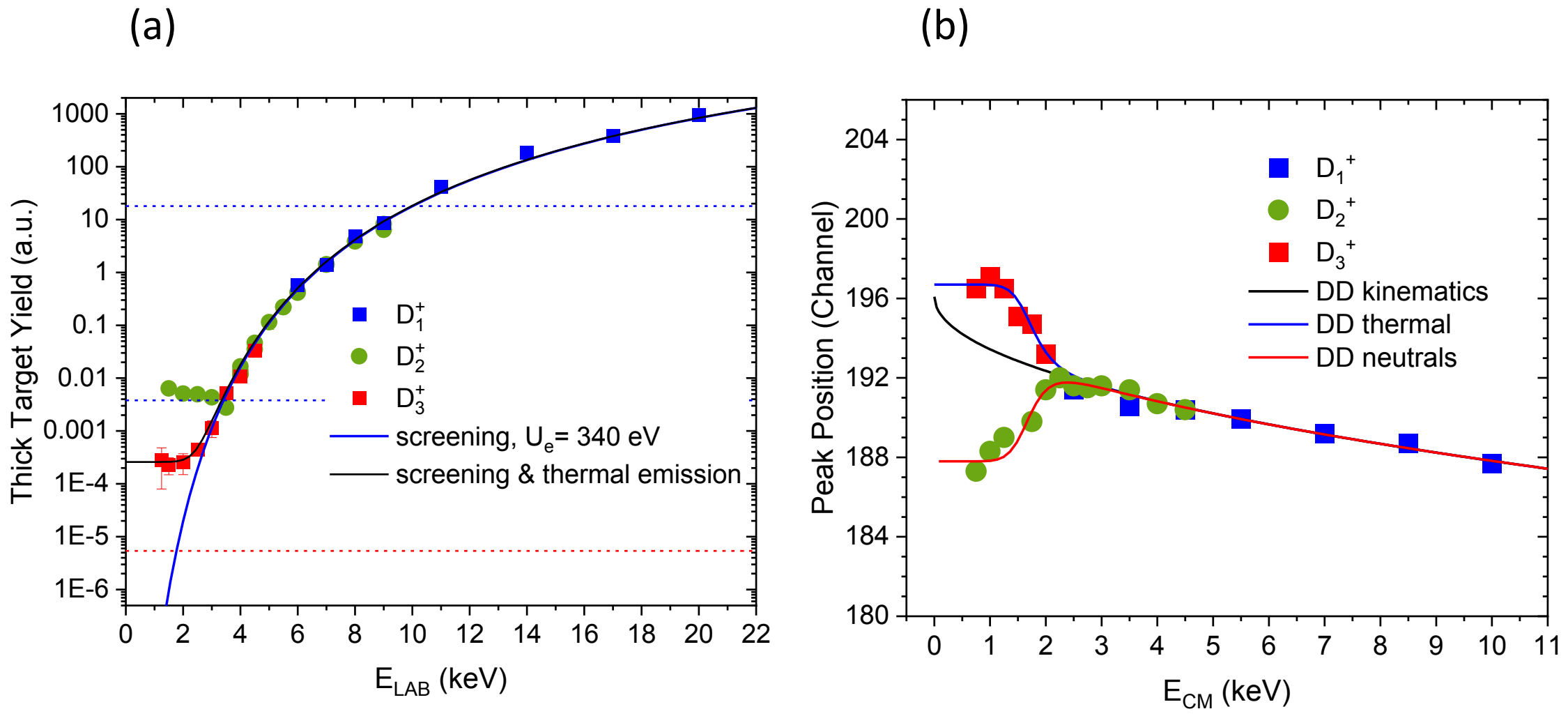

**Fig. 2: (a)** Thick target yield of the $^2H(d,p)^3H$ reaction measured on Zr. The blue full line corresponds to the theoretical screening curve (Eq. 3) obtained for the screening energy $U_e$=340±30 eV. The upper horizontal green dashed line shows the yield value measured with $D_2^+$ beam at $E_{LAB}$=10 keV per deuteron without using deceleration system. The lower green dashed line represents the yield value measured in the plateau region using the deceleration system, resulting from the neutralized beam fraction. The red dashed line corresponds to the expected yield for the neutralized $D_3^+$ beam compared to the value obtained at $E_{LAB}$=4 keV per deuteron without deceleration. The error bars of experimental points represent statistical uncertainty only. The systematic uncertainty due to deuteron density in the target is of about 10%. **(b)** Channel number of the proton peak position. At the lowest deuteron energies, the experimental data obtained with the $D_3^+$ beam can be described by a sum of thermal emission ($E_{CMS}$=0) and the fast-decreasing direct component corresponding to the screening curve from Fig. 2a (black curve). The proton peak position obtained for the $D_2^+$ beam can be achieved as a sum of the direct component and a constant neutral beam contribution. High energy points follow the energies resulting from the kinematics of the $^2H(d,p)^3H$ reaction (blue curve, see the Methods section).

The expression above can be applied when the astrophysical S-factor is a weakly energy-dependent function. In our case, it is a sum of two components: a constant value $S_0 =$ 57 keV b resulting from the known broad resonances in the compound nucleus $^4$He [27] and the $0^+$ threshold resonance $S_R$ [16]. This resonance has a pure deuteron-deuteron structure. Therefore, the proton and neutron channels are strongly suppressed, and the electromagnetic internal positron-electron pair creation dominates at the lowest deuteron energies.

The measured thick target yields can be also compared to a theoretical one by calculation of a corresponding ratio, determined for a fixed value of the screening energy:

$$F(E) = Y_{\exp}(E)/Y_{\mathrm{scr}}(E, U_e) \tag{4}$$

Most often, this so-called enhancement factor is estimated for $U_e = 0$. However, here we use it in conjunction with a fitted screening energy to more easily estimate the resonant contribution to the experimental yield values.

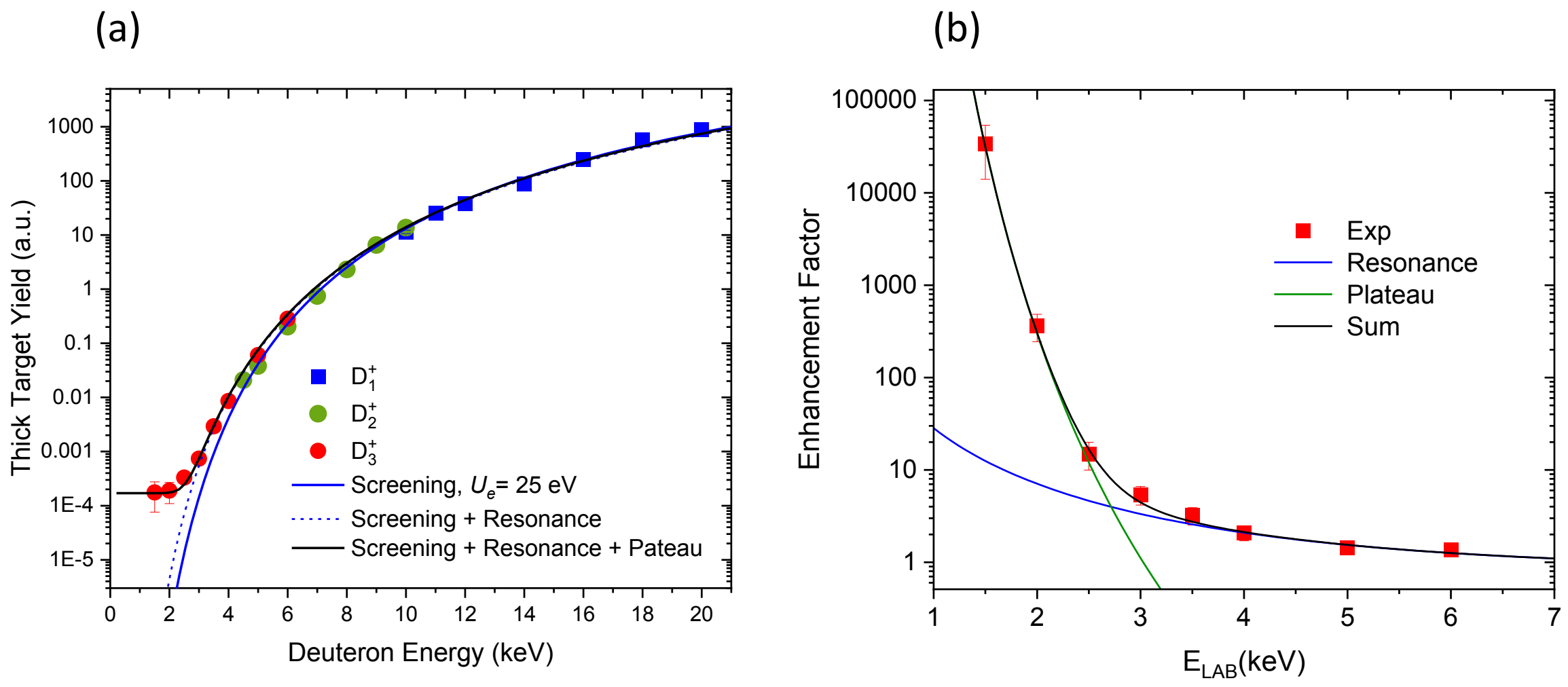

**Fig. 3: (a)** Thick target yield of the $^2$H(d,p)$^3$H reaction measured on Ti. The black curve represents a sum of the direct screening component and the thermal one determined by the constant yield plateau. The full blue curve results from the direct yield component calculated with $U_e$=25 eV. The dashed blue line takes into account the excitation of the threshold resonance. **(b)** Enhancement factor determined for the $^2$H(d,p)$^3$H reaction on Ti as a ratio of the measured thick target yield and the values calculated with the screening energy $U_e$=25 eV (Eq. 4) The blue curve presents the enhancement due to threshold resonance, and the green one corresponds to enhancement in the plateau energy region.

The experimental thick target yields determined for all studied target materials (Fig. 2-4) decrease over the measured energy range by many orders magnitude following a theoretical screening curve (Eq. 3) assuming a constant astrophysical S-factor of 57 keV barn. The “screening curves” were fitted using two parameters: the normalization factor and the screening energy. The corresponding values are presented in Table 2. Only in the case of Ti, for which the experimentally determined screening energy is very low, the small threshold resonance contribution could be unambiguously determined and explicitly added to the theoretical S-factor. The resulting enhancement factors is depicted in Fig. 3b. The physical parameters of the studied target materials are collected in Table 1.

The experimental yields were determined as the number of proton events divided by collection time in thousands of seconds and the deuteron-particle current expressed in µA. Data obtained for deuterated Zr (Fig. 2) can serve as a benchmark, as this target is highly stable and was extensively studied in the past through the use of advanced diagnostic techniques such as positron annihilation spectroscopy (PAS), Auger electron spectroscopy (AES), X-ray diffraction (XRD), and scanning electron microscopy (SEM) [12-14]. This special role of Zr applies particularly to the determination of the screening energy, the yield plateau and the energy shift of the proton line. Proton yield measurements could be performed down to a deuteron energy of 1.25 keV corresponding to a center-mass-energy of 0.625 keV. The fitted screening energy of 340$\pm$30 eV is much higher than that obtained for a defect free Zr target [15] of about 100 eV, which is in agreement with theoretical predictions based on the self-consistent dielectric function theory [10]. It was shown that crystal lattice defects induced by deuterium implantation and a small contamination by oxygen atoms can increase the screening energy for Zr to as much as 500 eV [13].

The experimental thick target yield determined for the Ti target using atomic and molecular deuterium beams is presented in Fig. 3. The fitted "screening curve" delivers a very high normalization factor, comparable to Zr, which confirms a high deuteron density achieved for the target. The experimental yields measured for deuteron energies below 6 keV are significantly higher than the theoretical curve obtained with the screening energy $U_e$=25$\pm$10 eV (the full blue curve), which matches the higher energy points very well. This yield enhancement observed for energies down to 3.5 keV could be explained with the contribution of the $0^+$ threshold resonance (the blue dotted line). We have used here the same proton resonance width of 40 meV and phase shift of 115°, which were previously obtained in our experiment on Zr [15]. Below the deuteron energy of 3.5 keV, the influence of the yield plateau is visible, value of which was estimated to be of 1.4x$10^{-4}$, being almost twice lower than for Zr. Thus, the yield values in the plateau region are subject to relatively higher statistical uncertainties. The enhancement factor defined as a ratio between the experimental thick target yield and the theoretical value expected for the screening energy $U_e$=25 eV reaches five orders of magnitude and is shown in Fig. 3b. Due to the relatively small screening energy of Ti, the resonance contribution to the experimental yield can be recognized very easily. It leads to the enhancement factor of 10 at the lowest measured beam energy.

The experimental data achieved for the Pd target are presented in Fig. 4. The screening energy of 400$\pm$40 eV was fitted similarly to that of Zr without taking into account the resonance contribution, which could be neglected. The thick target yield values obtained at higher deuteron energies and the corresponding normalization factor point to about 30 times lower deuteron density in Pd compared to that in Ti, giving the stoichiometric ratio D:M=0.05:1. At deuteron energies below 6 keV, a yield plateau can be observed, the value of which is 3 orders of magnitude higher than that determined for Zr using the same normalization factor for both screening curves. For comparison, the yield plateau was reached at deuteron energy 3 keV for Zr and 2 keV for the Ti target.

Measurements for Pd could be performed at higher deuteron beam energies leading to much better proton detection statistics. In addition, a relatively thin aluminum foil (1 µm) placed in front of the silicon detector was used, which allowed the separation of two different components of the proton line (see Fig. 1) at energies below 6 keV. The lower one follows the reaction dynamics, and its amplitude strongly decreases with the lowering projectile energy. The component at higher proton energy does not significantly change the count rate measured and can be observed at the same channel, independent of the beam energy. This component dominates the strength of the proton line in the yield plateau region and can be identified as a result of the thermal DD fusion.

### 2.2 Reaction yield plateau

At the lowest beam energies, the thick target yield does not follow "the screening curve" of Eq. 3 – a constant plateau of the reaction yield can be observed (see Fig. 2-4), which results in very high enhancement factors. The lowest yield plateau has been measured using the $D_3^+$ beam at the deuteron beam energies below 2.0 keV for the Ti target, which corresponds to the cross section of about $50\pm10$ pbarn. The effect of the yield plateau is clearly correlated with the higher energies of emitted protons than those based on nuclear reaction kinematics. The detected proton energies in the plateau region reach the maximum values corresponding to the emission of 3 MeV protons from a resting center mass system. This indicates that we observe DD fusion at thermal energies on the order of tens of meV.

The opposite effect – a reduction in the energy of the emitted protons – can be observed due to the beam neutralization effect when relatively high voltages are applied to the deceleration lenses to significantly lower the beam energy. In Fig. 2, the measured thick target yield on Zr using $D_2^+$ beam together with the detected proton energies are also presented. The ion source potential was set at 20 kV, and the deceleration voltage was gradually increased to values exceeding 18 kV to achieve the lowest beam energies. However, the deceleration lenses affected only charged ions, not neutral atoms. Such atoms can be formed through electron-exchange reactions with residual deuterium gas along the 1.5 m ion-beam section between the electric steering elements and the deceleration lenses (see Fig. 6 in the Methods section). The neutral beam contribution leads to an energy reduction of emitted protons since the neutral atoms have much higher energies stemming solely from the ion source voltage. The absolute value of the neutral beam fraction can be determined as a ratio of the reaction yield measured at 20 keV (without deceleration) and the yield plateau value produced by high energy neutrals at low beam energies. The neutral beam fraction could be experimentally determined for $D_1^+$ and $D_2^+$ beams, leading to a similar value of about $3x10^{-4}$. This number agrees with a theoretical estimation assuming the known neutralization cross section [28-30] of about $10^{-15}$ cm$^2$ for all deuteron beam species on the $D_2$ rest gas and the vacuum pressure of about $8x10^{-9}$ mbar in this part of beam line. Thus, to reduce the neutral contribution to the experimental reaction yield at the lowest energies, we used $D_3^+$ beam and the ion source potential of 12 kV, which leads to the yield values induced by neutrals being almost two orders of magnitude lower than the observed plateau.

In the case of measurements on Pd, the experimentally determined yield plateau is much higher than for Zr and Ti, and therefore the use of the deceleration system was not necessary. Thus, there was no beam component of high energy neutral atoms, which could influence the data obtained. The average position of the proton peak depending on the energy of the deuteron beam in the center of mass system is shown in Fig. 4b. Like the results obtained for Zr, we observe that the protons detected in the plateau energy region are emitted from the resting frame with $E_{CM}$=0. Analogously, the red curve in Fig. 4b is calculated as a weighted average of the peak position resulting from the reaction kinematics (corresponding to the screening curve for $U_e$=400 eV) and the constant value representing the plateau yield contribution.

The dependence of the proton peak position on the deuteron beam energy provides a decisive argument regarding the mechanism of the $^2H(d,p)^3H$ reaction. We can easily distinguish between the thermal emission of 3 MeV protons, protons emitted via the direct DD fusion process (governed by standard reaction kinematics), and protons produced by high-energy neutral atoms (Fig. 1b).

**2.3 Thermal reaction rates**

High energies of detected protons in the yield plateau region corresponding to emission from the resting center mass system can be explained by formation of ion tracks by the deuteron beam. According to the thermal spike model, in a small cylindrical region around to the primary collision cascade of impinging projectiles, the temperature is strongly increased, even above the melting point of the studied materials. After the beam deuteron is stopped, over the next few tens of picoseconds the temperature drops and recrystallization occurs, along with the expansion of the volume of the ion track. The process can be described within the thermal spike model with the heat diffusion equation for spherical coordinates:

$$C_a \frac{\partial T(r.t)}{\partial t} = \frac{1}{r}\frac{\partial}{\partial r}[rK_a\, \partial T(r,t)/\partial r] + A_a(r,t) + gA_e(r,t) \qquad (5)$$

where $C_a$ and $K_a$ are the crystal lattice heat capacity and thermal conductivity of the irradiated material, respectively. The energy absorbed within the ion track is a sum of the energy lost due to the atomic collisions and electronic excitation of the target atoms. The latter increases the temperature of the primary ion track to a fraction $g$, which arises from electron-phonon coupling for the studied material. The sum of both contributions results in the total absorbed energy in the crystal lattice $\epsilon$, which is related to the ion beam energy. The heat diffusion equation can be solved analytically for the spherical ion track and the target ambient temperature $T = 0$°K [21]:

$$T(r,t) = \frac{\epsilon C_a^{1/2}}{(4\pi K_a t)^{3/2}} exp\left(-\frac{C_a r^2}{4K_a t}\right) \tag{6}$$

leading to a Gaussian temperature profile whose width increases with time, accompanied by a decreasing maximum value. Stopping the projectiles in the target takes about 0.1 ps, which is also needed to reach a well-defined temperature. For the $ZrD_2$ target, it can be shown that the total energy absorbed within an ion track remains almost constant for deuteron beam energies of several keV if about 10% of the energy absorbed by electrons is transferred to the crystal lattice [31]. Considering Eq.6, the temperature at the center of the ion track in the $ZrD_2$ target will reach 2910°K, and the diameter of the ion track, defined as the distance at which the temperature drops to half its central value, will be about 1 nm. For Ti and Pd targets, these values are very similar, as their heat capacities and thermal conductivities are close to those of Zr (see Table 1). Since the actual ambient temperature of ion tracks is about 300°K and the ion track expands not in the full $4\pi$ space but in half, its temperature will be considerably higher and its lifetime will be longer, which causes local melting of the target material. However, the main effect that determines the lifetime of the thermal spike, ranging between 100 and 1000 ps [25,26], is the recrystallization of the molten tracks that generates additional heat and intensifies diffusion. More accurate values require numerical modeling. Therefore, the approach presented above only illustrates which parameters of the target material determine the spatial and temporal expansion of ion tracks. Since the heat capacities and thermal conductivities and their ratio for all metallic materials studied here are very similar, we can conclude that the temperature profiles will be also very similar.

Nevertheless, to determine the absolute nuclear reaction rate arising from the number of deuteron-deuteron collisions within the ion track, it is necessary to estimate its average dimensions and temperature. For that, we can apply a simple model based on the phonon density induced by projectiles in the target material due to the elastic collision cascade. Since the total energy absorbed in the lattice (including electron excitation transferred to the lattice by the electron-phonon coupling) is known, and the energy of acoustic phonons in ZrD2 is about 100 meV, the initial phonon density of 280 phonons/nm$^3$ corresponding to the temperature of 2540°K can be obtained [31] in agreement with the thermal spike estimation. It is assumed here that the initial ion track volume for 2 keV deuterons in ZrD2 is about 21 nm$^3$ as a result of 1nm$^2$ of the lateral area and 21 nm of the range of beam deuterons [32]. The mean value of the phonon density (a half of the initial density) can be then used to determine the average ion track temperature applying the Bose-Einstein statistics of phonons. For $ZrD_2$, this model leads to the average ion track temperature of 1080°K and the volume of about 1870 nm$^3$.

The increased temperature is responsible for an enhanced deuteron diffusion and free moving deuterons in the ion tracks, which can be expressed by the Arrhenius relation:

$$N = N_0 \exp\left(-E_A/kT\right) \tag{7}$$

where $E_A$ is the activation energy and $N_0$ stays for the deuteron density in the target. Consequently, the reaction rate of the $^2$H(d,p)$^3$H reaction could be calculated assuming Maxwell-Boltzmann velocity distribution:

$$\mathcal{R}(T) = N_0 N \langle \sigma\, v \rangle = N_0 N \,\frac{(8/\pi)^{1/2}}{\mu^{1/2}(kT)^{3/2}} \int_0^\infty \sigma(E)\, E\, exp(-E/kT)\, dE =$$

$$= N_0 N \,\frac{8\pi^{3/2}\hbar^3 E_R^{1/2}}{\mu^2\, a\, (kT)^{3/2}}\, (E_G/U_e)^{1/2}\, exp\left[-(E_G/U_e)^{1/2}\right]\, exp(-E_R/kT)\, \frac{\Gamma_p}{\Gamma} \tag{8}$$

The formula above was derived under assumption that the single particle threshold resonance has a very small width (< 1 eV) and therefore the standard reaction rate estimation used in nuclear astrophysics for narrow resonances [33] can be utilized. Accordingly, the reaction rate in the yield plateau region is determined by three most important factors: the increased deuteron diffusion determined by the density of moving deuterons $N$, the electron screening effect given by the screening energy $U_e$ and the enhancement due to the threshold resonance described by its resonance energy and widths.

## 3. Comparison of thermal fusion in different materials

The observation of the yield plateau and the spectral component correlated with the proton emission from the resting center mass system let us conclude that the protons are produced by DD fusion occurring at thermal energies. The corresponding nuclear reaction rate can be calculated using Eq. 8, taking into account material specific quantities. All the studied metallic targets Zr, Ti and Pd are characterized by different values of physical parameters which allows to test the ion track model of the thermal DD fusion. It assumes that the atomic collision cascade induced by projectiles locally increases crystal lattice temperature leading to enhanced deuterium diffusion and nuclear reaction rates. The experimentally observed count rate of thermal protons can be estimated using the nuclear reaction rate $\mathcal{R}(T)$ and the average lifetime $\tau$ and volume $V$ of ion tracks:

$$\mathcal{N}(T) = I \cdot V \cdot \tau \cdot \frac{\Omega}{4\pi} \cdot \mathcal{R}(T) \tag{9}$$

where in the case of Zr, the projectile current $I = 90$ µA at deuteron energy 2 keV defines the number of tracks per second and $\Omega$ stays for the solid angle of the detector. For the average track temperature of 1080°K, track volume of 1870 nm$^3$, screening energy 340 eV, ion track lifetime of 100 ps and $\frac{\Omega}{4\pi}$ =0.0032, we obtain a very conservative count rate estimate of one proton per 100,000 s. This value is merely 3 times lower than that observed in our experiment within the energy range of the yield plateau.

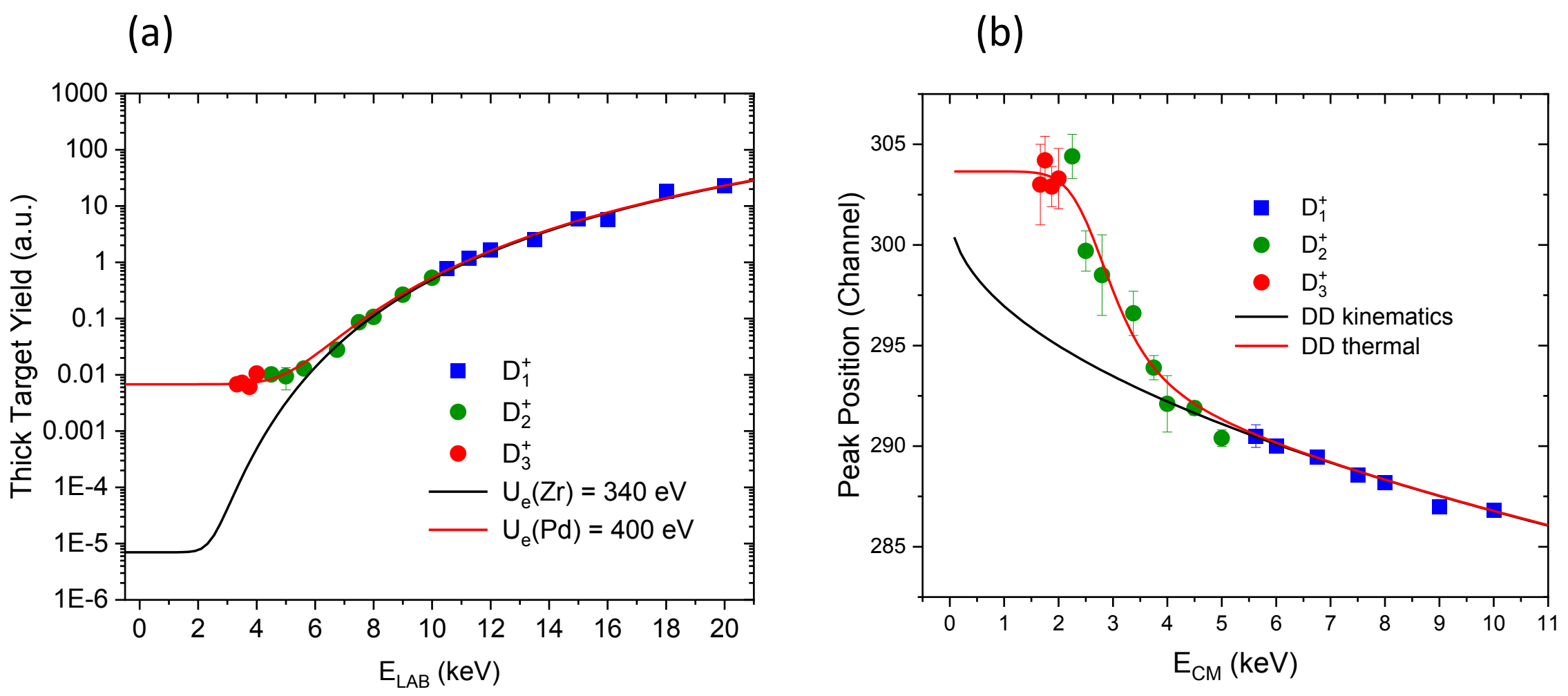

**Fig. 4: (a)** Thick target yield of the $^2H(d,p)^3H$ reaction in Pd. The red curve represents a sum of the direct screening component and the thermal one determined by the constant yield plateau. For a comparison, the yield curve obtained for Zr, normalized to high energy points, is also presented (black curve). The yield plateau level of Pd is 3 orders of magnitude higher than that for Zr. **(b):** Channel number of the average proton peak position. The red curve represents a sum of both spectral contributions: direct screening component and thermal one. The proton energies measured in the yield plateau region are significantly larger than those expected due to reaction kinematics (black curve).

The uncertainty in this estimate is relatively large and amounts to one to two orders of magnitude, mainly due to the uncertainty of the screening energy and electron-phonon coupling. However, these uncertainties should be considered in relation to the reaction rate which is about 20 orders of magnitude higher due to the electron screening effect and threshold resonance when compared to the value obtained for the defect free Zr target ($U_e$=100 eV). Similarly, the average ion track temperature of 1080°K leads to the increased deuteron diffusion in comparison to room temperature and results in the enhancement factor of the fusion rate by more than 5 orders of magnitude. This number also shows how important the high temperature within ion tracks is for thermal proton emission.

Considering the uncertainties of absolute value of the plateau yield, it is convincing that a relative comparison between the values obtained for different target materials is more appropriate. We assume here that the ion track lifetime and the average temperature of all studied materials should be similar due to very similar ratios between heat capacity and thermal conduction coefficients (see Table 1). On the other hand, the activation energies governing deuterium diffusion in the analyzed materials differ significantly. Thus, this factor may play a crucial role in the observed yield plateau level. Especially, Pd is known for a strong hydrogen diffusion. Its activation energy of 150 meV [42] is smaller than for Zr ($E_A$=413 meV [40]), which results in the hopping deuteron number being **16.9** times greater for Pd than for Zr. Higher screening energy leads to an additional enhancement by a factor of **61**. Both effects combined result in the **1030** times higher yield plateau, which is very close to the experimental value (see Table 2 and Fig. 4). Considering that the deuteron density was in the Pd target **37** times smaller than in Zr, we get finally the absolute plateau yield of about **27** times greater for Pd than for Zr.

**Table 1.** Physical parameters of studied materials.

| Target | Mass Density (g/cm$^3$) | Thermal conductivity $K_a$ (W/cm/K) | Heat Capacity $C_a$ (J/cm$^3$/K) | Central Temperature* (K) | Melting Point** (K) |
|---|---|---|---|---|---|
| $ZrD_{1.74}$ ($\varepsilon$) | 5.61 | 0.11[34] | 2.59[37] | 2910 | 2120 |
| $TiD_{1.93}$ ($\delta$) | 3.97 | 0.09[35] | 2.83[38] | 4110 | 1940 |
| $PdD_{0.054}$ ($\alpha$) | 12.02 | 0.10[36] | 2.93[39] | 3570 | 1830 |

* Central temperature in the ion track calculated according to Eq. 6 for t=0.1 ps and $\epsilon$=0.58 keV (assuming the deuteron beam energy of 2 keV [31]).

** Melting points for pure metals.

In the case of the Ti target, the measured screening energy was much lower ($U_e$=25 eV) than achieved for other metals and corresponds to values, which are expected for insulating targets. This is because of a relatively strong embrittlement effect at the target surface at high hydrogen density. Similar small screening energies characteristic for insulating materials were determined for Ti in experiments of other groups [43,44]. Additionally, due to the strong hydrogen binding in Ti, the activation energy of 680 meV [41] is even higher than for Zr, leading to 17 times weaker deuterium diffusion at 1080°K. Surprisingly, the absolute yield plateau for Ti observed experimentally is, however, only two times lower than for Zr, whereas the deuteron density is comparable. This finding supports the thermal spike model, which considers emission of protons from high temperature, highly excited ion track regions. As already observed in experiments using swift heavy ions [45], the irradiated insulator materials can change their structure within ion tracks to a metallic-like phase. This leads automatically to a strong increase of the screening energy, which is characteristic for metals. Consequently, the cross section for DD fusion at thermal energies rises by more than twenty orders of magnitude, allowing for observation of the yield plateau. Accordingly, the screening energy of the hot Ti target should be around 370 eV to compensate the poor hydrogen diffusion.

**Table 2.** Experimentally determined screening energies and yield plateaus. The activation energy from Eq. 7 describes deuterium diffusion at high temperature. The yield normalization factor corresponds is proportional to the experimental deuteron density.

| Target | D/M Ratio | Activation Energy (meV) | Yield Normalization | Screening Energy (eV) | Yield Plateau Absolute* | Yield Plateau Relative** |
|---|---|---|---|---|---|---|
| Zr | 2 | 413[40] | 14,600,000 | $340\pm30$ | $2.6x10^{-4}$ | 1 |
| Ti | 2 | 680[41] | 12,690,000 | $25\pm10$ (370[#]) | $1.4x10^{-4}$ | 0.6 |
| Pd | 0.05 | 150[42] | 390,000 | $400\pm40$ | $6.9x10^{-3}$ | 1000 |

* Experimentally observed yield plateau.

** Yield plateau normalized on the deuteron density and the value determined for Zr.

[#] Screening energy for Ti, which is needed to explain the experimental yield plateau.

## 4. Discussion and conclusions

In this paper, we have presented for the first time thick-target yield measurements of the $^2H(d,p)^3H$ reaction on deuterated Zr, Ti and Pd targets down to the previously unavailable deuteron energy in CMS of 0.625 keV. The measurements were performed using a standard accelerator technique, involving the precise determination of beam energy with an accuracy of about 10 eV and the separation of the individual atomic and molecular beam components applying an analyzing magnet. Previously, deuteron beam energies on the order of 0.8 keV for DD fusion were achieved in glow discharge experiments [46], which allowed for the generation of high deuteron currents. However, the very wide spread of deuteron energies and the complex composition of the hydrogen plasma significantly hindered a reliable analysis of the DD fusion process [47]. The exponential like decrease of thick target yields observed in our experiments could be described by penetration probability through the Coulomb barrier being reduced by the screening energy dependent on the target material, which agree very well with values obtained by other research groups. In all cases, a reaction yield plateau at the lowest deuteron energies has been observed. In particular, the results obtained for Pd are very convincing since the yield plateau could be achieved at much higher projectile energies without using a special deceleration system. Based on the proton peak energy detected by the Si-detector at backward scattering angle, it could be concluded that the DD fusion occurs in the resting center mass system, which corresponds to thermal fusion on the meV energy scale. In the case of measurements for Pd, it was also possible to resolve the spectral contribution corresponding to the thermal and direct components of the fusion process within the proton line, revealing a distinct dependence of these components on projectile energy. The thermal component can be characterized by a count rate independent of the projectile energy and a constant energy of the emitted protons.

All three studied metallic targets, Zr, Ti and Pd are characterized by different values of physical parameters as deuterium diffusivity, heat conduction and thermal conductivities which allows to test the ion track model proposed here to explain thermal DD fusion. It assumes that the atomic collision cascade induced by projectiles locally increases crystal lattice temperature leading to enhanced deuterium diffusion and nuclear reaction rates. Based on Eq. 8, which describes the main factors responsible for thermal proton emission: the deuterium density, the electron screening effect and the excitation of the threshold resonance, we were able to estimate both the absolute count rate observed in the yield plateau region and its relative changes when comparing different target materials.

According to the analytical solution (Eq. 6) of the heat diffusion equation, the temperature in the center of the ion tracks amounts at least to 3000°K, significantly more than the melting points of the materials studied. These temperatures can be reached already 0.1 ps after the projectiles initiate collision cascades. Since metals have high thermal conductivity coefficients, cooling of the molten ion tracks should take, in principle, a relatively short time, a few tens of ps [48]. However, the resolidification process, which adds an additional crystallization energy of 0.1-0.2 eV per atom, extends the ion track lifetime to several hundred ps or even several ns, which was also shown in laser induced experiments [49,50]. Conservatively estimated parameter values of Eq. 9 can fully explain the observed yield plateaus in DD fusion. This can be also verified by using Eq. 8 and 9 to calculate the reaction reactivity $\langle \sigma\, \upsilon \rangle$ and cross section from the experimentally measured count rate without assuming any details about resonance and screening energy. Correspondingly, the reactivity amounts to $1.6 \times 10^{-34}$ cm$^3$/s and the reaction cross section is about $5.3 \times 10^{-15}$ b. The latter can be estimated considering that average temperature within ion tracks is 1080°K. The result agrees very well with the previous cross section calculations based only on higher energy data [16].

One of the most important consequences of the presented experiments is also the confirmation of the existence of the threshold resonance in $^4$He. This very narrow single-particle $0^+$ resonance with the total width of about 200-300 meV increases the thermal reaction rate by about seven orders of magnitude and was recently confirmed by observation of emission of e+e- pairs [19]. In the case of measurements on Ti, the resonance contribution could be clearly observed also at higher deuteron energies due to the very small screening energy. The enhancement factor of the $^2$H(d,p)$^3$H reaction compared to the bare nuclei cross section, neglecting the threshold resonance determined at the $E_d$=1.5 keV is larger than $10^5$.

Concluding, the achieved results confirm that DD reactions can take place at thermal energies due to high electron screening energies in the hosting metallic environments and the threshold resonance observed previously. The experimental method employed here will likely allow to determine the reaction rates of thermal DD fusion more precisely if the material parameters will be known much more exactly and numerical calculations using for instance molecular dynamics approach will be applied. This may be particularly important for determining energy production in the interiors of brown dwarfs and giant planets. On the other hand, identifying the main factors influencing the measured reaction rates makes it easier to find special metallic materials that would produce the highest amount of energy in terrestrial laboratories. This opens new perspectives for future materials research aimed at developing a new fusion energy source.

After submission of this manuscript, also an American research group reported an independent observation of a yield plateau in the $^2H(d,p)^3H$ reaction for deuterated Pd and Ti targets [51]. The yield plateau level achieved for Pd agrees with our data and is, however, much larger in the case of Ti. The relatively big energy uncertainty of the ion source used and the absence of a beam analyzing magnet could lead to significantly overestimated screening energies. Therefore, any energy shift analysis of emitted protons could not be performed, which resulted in a different interpretation of the obtained data.

## 5. Methods

### 5.1 Experimental setup

The experiments were conducted at the eLBRUS Ultra High Vacuum Accelerator of the University of Szczecin, Poland. The facility combines a very good vacuum of $10^{-10}$ mbar in the target chamber with a high current ion beam up to 1 mA on the target [52]. Due to dropping cross sections of nuclear reactions at very low energies below the Coulomb barrier, an important feature of the accelerated ion beams is their very good energy definition of 10 eV and the long-term stability of about 5 eV, provided by the ECR ion source. The electric voltage put on the ion source with respect to the grounded target chamber gave directly the energy of the accelerated charged particles. The same beam energy accuracy and the selection of a Q/M beam components were ensured by a 90° double focusing analyzing magnet equipped with a high precision Hall probe for B-field measurement. The accelerated system has been recently upgraded (see Fig. 6) and put on a special platform increasing or decreasing its potential compared to the target chamber at the ground potential. In front of the target chamber a decelerating system consisting of two electric lenses has been additionally installed. Thanks to the modification, the ion source could be operated at high potentials delivering high currents independent of the final beam energies. After deceleration, a beam current of several hundred microamps at the lowest projectile energies of about 1 keV, focused to a 7 mm diameter beam spot on the target could be achieved. The highest acceleration potential is 26 kV. For the measurements performed on Pd target, the deceleration unit was not used since the yield plateau could be observed at higher beam energies. The beam current stability measured on the target was in all cases better than 0.5%. Secondary electron emission from the target increased the ion beam charge collected only by 10-15% for the entire energy range of the used deuteron beam, which was checked by positively biasing the target holder with a voltage up to 60 V.

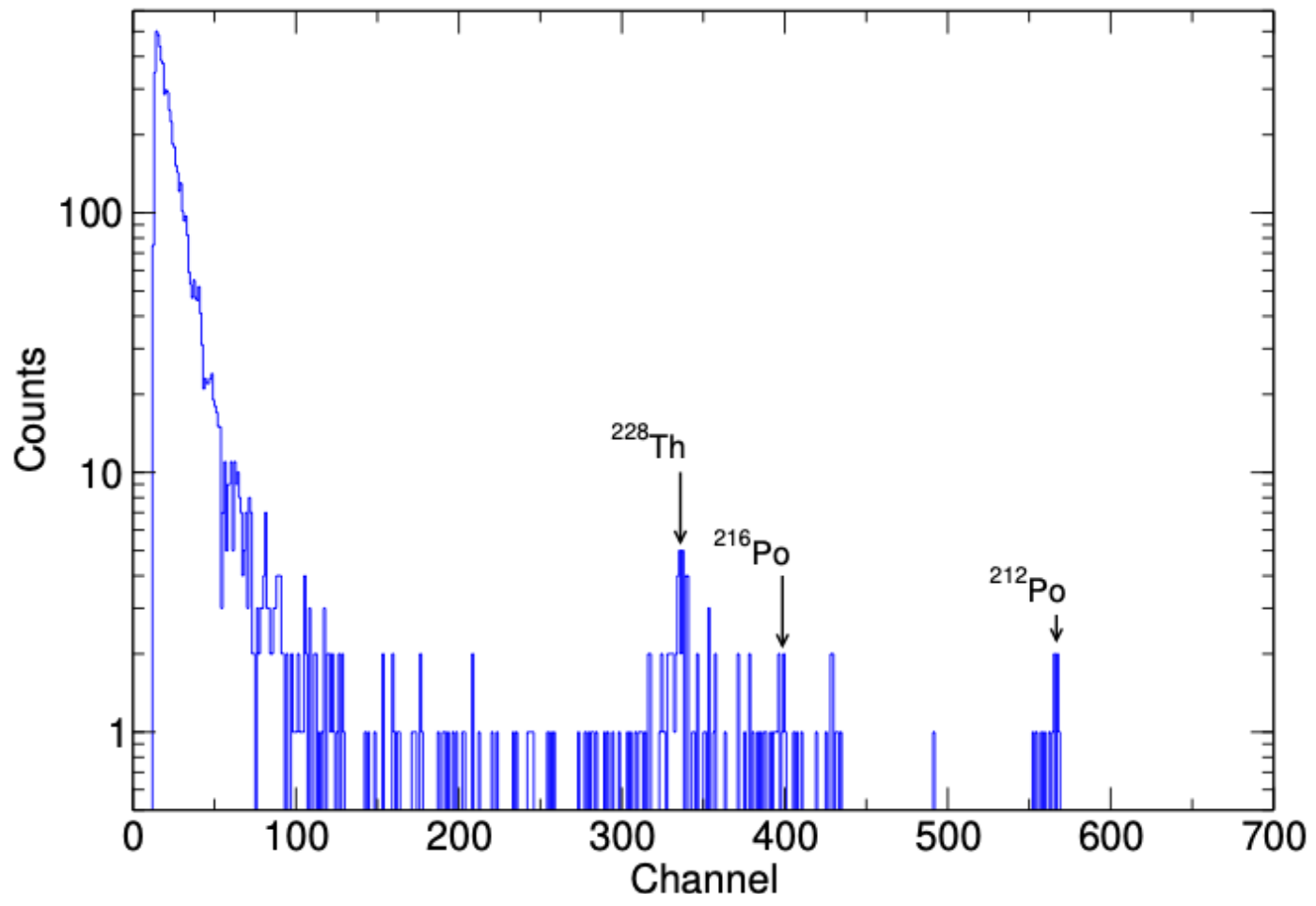

**Fig. 5:** Natural background spectrum characteristic for the $^{232}$Th alpha-decay series: $^{228}$Th 5.52 MeV, $^{216}$Po 6.91 MeV, $^{212}$Po 8.95 MeV. The energies measured are lower than the nominal decay values due to the energy loss of alpha particles in the protection foil in front of the Si detector. The spectrum was measured for 5 days.

In the present experiment, atomic and molecular deuteron beams $D_1^+$, $D_2^+$ and $D_3^+$, analyzed magnetically were applied. Charged particles emitted in D-D reactions — $^3$He nuclei (0.8 MeV), $^3$H nuclei (1.02 MeV), and protons (3 MeV) — were detected using a low-noise PIPS silicon detector (100 µm thick with an active area of 100 mm² for Zr and Ti targets), positioned 5 cm from the target at an angle of 135° relative to the beam. For measurements involving Pd, a detector with greater thickness (1000 µm) and an active area of 150 mm² was used, placed at a distance of 6 cm, to also allow for the detection of the internal $e^+e^-$ pair creation process. In front of the detector, 1 µm (for Pd) and 4 µm (for Zr and Ti) thick aluminum foils were located to prevent elastically scattered deuterons from entering the detector. The energy resolution of both detectors was about 20 keV. The use of a thinner aluminum foil for Pd enabled an improvement in overall energy resolution and the separation of the two components of the proton spectral line (see Fig. 1). As targets, 1mm thick Zr, $Ti_{0.993}Pd_{0.007}$ and Pd plates implanted to the highest possible stoichiometry $ZrD_2$ was applied. Deuterated Zr targets are known for their high stability and isotropic deuterium density distribution with small fluctuations of about 10%, as well as for their high electron screening energies [15]. The target temperature was kept below 45°C by reducing the beam current at higher deuteron energies and using a special air-cooling system. A similar setup was also used for the deuterated Ti target, which remained stable even at higher beam intensities. In the case of Pd, only a low deuterium stoichiometry (around 5%) could be achieved due to the weak hydrogen bonding in this metal. At higher beam intensities, the deuterium density decreased. Consequently, the beam power was maintained at a constant level of approximately 0.6 W for all energy points, resulting in a target temperature of about 30-32°C and fluctuation of the deuteron density of about 15%. The target stability was proved by means of repeated proton yield measurements performed at the same energies (both low and higher energies) during the whole experimental campaign. There were no signs of any target degradation, the count rate remained stable. The experimental setup was presented in detail together with Monte Carlo simulations of particle spectra in [18].

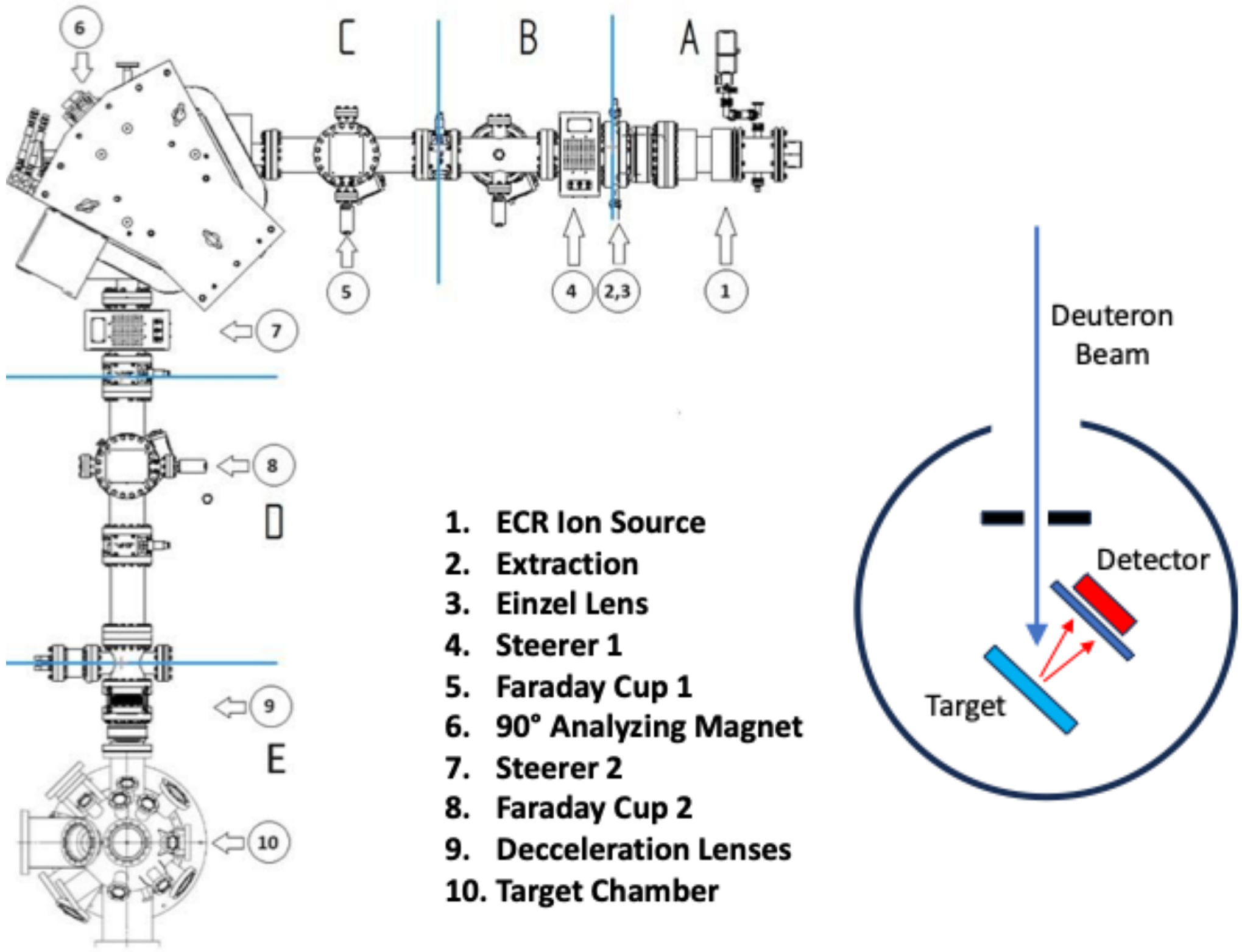

**Fig. 6:** Ultra-high vacuum accelerator system at the University of Szczecin, Poland. The displayed sections of the system correspond to different units of the differential pumping system. The right side shows the position of the detector in the target chamber relative to the deuteron beam.

Experimental spectra measured at different energies using atomic and molecular deuterium beams are displayed in Fig. 1. At the lowest beam energies, the spectra were collected for more than one week. A corresponding background in the energy region of the proton line of a few counts was subtracted. The stability of the energy calibration of the Si detector was repeatedly checked during the measurements by alternating short measurements at higher deuteron energies with measurements at lower energies, requiring longer measurement times. Additionally, at the lowest beam energies, energy calibration was also assured by controlling the line position of the natural background dominated by three main alpha-decays (see Fig. 5).

### 5.2 Thick target yield vs. reaction rate

The screened reaction cross-section below the Coulomb barrier can generally be expressed by the astrophysical S-factor:

$$\sigma_{scr}(E) = \frac{1}{\sqrt{E(E+U_e)}}\, S(E)\, exp\left(-\sqrt{\frac{E_G}{E+U_e}}\right) \tag{10}$$

where the astrophysical S-factor $S(E)$ in absence of narrow resonances is usually a slowly changing function of the CM energy. In the low energy accelerator studies, the cross section cannot be determined directly as the range of projectiles is very low and the cross-section changes strongly. Instead, the thick target yield being a sum of the reaction contributions from different target depth is a standard quantity that can be obtained:

$$Y_{scr}(E) = N_0 \int_0^R \sigma_{scr}(E)\, dx = N_0 \int_0^E \frac{\sigma_{scr}(E)}{|dE/dx|}\, dE \tag{11}$$

where the density of reacting nuclei in the target $N_0$ is assumed to be constant along of the pathway of projectiles up to their range $R$ . The above expression can be compared with the experimental value, which, for practical reasons, have been defined as the number of detected protons divided by the deuteron beam current (expressed in μA) and the measurement time (in thousands of seconds). The stopping power function $|dE/dx|$ has to components: electronic and nuclear. The electronic part describes energy loss mechanisms related to atomic excitation and ionization which is proportional to $\sqrt{E}$ in the low energy limit. For projectile energies below 1 keV, the nuclear stopping power associated with scattering on lattice atoms begins to dominate. Since the yield function decreases exponentially with lowering energies and our experiments are performed for deuteron energies higher than 1 keV, we can approximate the stopping power value as $|dE/dx| = C\sqrt{E}$ , which leads to an expression:

$$Y_{scr}(E) = N_0 \int_0^E \frac{\sigma_{scr}(E)}{C\sqrt{E}}\, dE = \frac{N_0}{C}\int_0^E \frac{\frac{1}{\sqrt{E(E+U_e)}} S(E) \exp\left(-\sqrt{\frac{E_G}{E+U_e}}\right)}{\sqrt{E}}\, dE \tag{12}$$

This integral can be calculated analytically:

$$Y_{scr}(E) = \frac{2N_0\, S(E)}{C\sqrt{E_G}}\left[\exp\left(-\sqrt{\frac{E_G}{E+U_e}}\right) - \exp\left(-\sqrt{\frac{E_G}{U_e}}\right)\right] \simeq \frac{2N\, S(E)}{C\sqrt{E_G}} \exp\left(-\sqrt{\frac{E_G}{E+U_e}}\right) \tag{13}$$

assuming that the astrophysical S-factor does not change strongly and that $E \gg U_e$. The S-factor is a coherent sum of a broad resonance contribution $S_B$ and the narrow $0^+$ threshold resonance [15,16] $S_R$:

$$S(E) = S_B + S_R(E) + 2\left(\frac{1}{3} S_B S_R(E)\right)^{1/2} cos(\varphi_B^{0+}) \tag{14}$$

where $S_B$= 57 keV b is taken as a constant [27], and only one third of its value contributes to the interference effect. The resonance component can be described by a Breit-Wigner formula:

$$\sigma_R(E) = \frac{\pi}{k^2}\frac{\Gamma_d\Gamma_p}{(E-E_R)^2+\Gamma^2/4} \simeq \frac{\pi}{k^2}\frac{\Gamma_d\Gamma_p}{E^2} \tag{15}$$

Here we consider $E_R \to 0$. The deuteron partial width $\Gamma_d$ takes the value for the single-particle resonance:

$$\Gamma_d(E+U_e) = 2kaP(E+U_e)\frac{\hbar^2}{\mu a^2} = 2k\frac{\hbar^2}{\mu a}(E_G/(E+U_e))^{1/2}\, exp\left[-(E_G/(E+U_e))^{1/2}\right] \tag{16}$$

and

$$S_R(E) = \frac{(2E_G)^{1/2}\hbar^3}{a\mu^{3/2}}\frac{\Gamma_p}{E^2} \tag{17}$$

where $\Gamma_p$ is the proton partial resonance width, and $a$ and $\mu$ stay for the DD channel radius of 7 fm and the reduced mass, respectively. Since the resonance contribution to the thick target yield in the energy range studied is relatively low compared to a strong screening enhancement, the screening curves presented in Fig. 2 and 4 do not include resonance contribution and the S-factor was assumed to be constant. Only in the case of the Ti target with a very low screening energy, the threshold resonance was considered (Fig. 3).

Oppositely to the standard accelerator experiments, where a projectile reacts with resting target nuclei, in the case of the classical plasma at temperature $kT$, where all the constituents interact with each other, the nuclear reaction rate is a proper quantity to describe the observed number of reactions:

$$\mathcal{R}(T) = \frac{N^2}{2}\langle\sigma\, v\rangle = \frac{N^2}{2}\frac{(8/\pi)^{1/2}}{\mu^{1/2}(kT)^{3/2}}\int_0^\infty \sigma(E)\, E\, exp(-E/kT)\, dE \tag{18}$$

Here we assumed the classical Maxwell-Boltzmann velocity distribution. In the case of a narrow Breit-Wigner resonance, considering that its resonance strength is distributed over a small thermal energy spectrum, we get:

$$\mathcal{R}(T) = \frac{N^2}{2}\left(\frac{2\pi}{\mu\, kT}\right)^{3/2}\hbar^2\, exp(-E_R/kT)\,\frac{\Gamma_d\Gamma_p}{\Gamma} \tag{19}$$

The expression for the deuteron single-particle resonance width (Eq.13) at $E = 0$ finally leads to:

$$\mathcal{R}(T) = \frac{N^2}{2} \frac{8\pi^{3/2}\hbar^3 E^{1/2}}{\mu^2 \, a \, (kT)^{3/2}} \, (E_G/U_e)^{1/2} \, exp\left[-(E_G/U_e)^{1/2}\right] \, exp(-E_R/kT) \, \frac{\Gamma_p}{\Gamma} \tag{20}$$

The screening function $exp\left[-(E_G/U_e)^{1/2}\right]$ is the main factor deciding on the absolute value of the reaction rate. However, the number density $N$ strongly depends on the temperature, as well. Its value can be correlated to the diffusion coefficient and can be expressed as $N = N_0 \, exp(-E_A/kT)$, where $E_A$ denotes the activation energy, equal to 413 meV for $ZrD_2$ [40]. The local temperature within the ion track induced by bombarding deuterons is related to a number of excited phonons described by the well-known nuclear stopping power function [32] and reaches an average value of about 1080°K. This corresponds an increase of the number of hopping deuterons in the crystal lattice, responsible for the DD collisions, by five orders of magnitude compared to room temperature. Its absolute value is, however, still relatively low $N/N_0 = 0.012$ . Therefore, in this case, the number of interacting DD pairs will not be given by a factor $N^2/2$ valid for classical plasma (see Eq. 18), but rather by the product $N_0 N$, which assumes that most of the deuterons will remain trapped in the lattice. This leads to the final formula for the reaction rate:

$$\mathcal{R}(T) = N_0 N \, \frac{8\pi^{3/2}\hbar^3 E^{1/2}}{\mu^2 \, a \, (kT)^{3/2}} \, (E_G/U_e)^{1/2} \, exp\left[-(E_G/U_e)^{1/2}\right] \, exp(-E_R/kT) \, \frac{\Gamma_p}{\Gamma} \tag{21}$$

### 5.3 Experimental thick target yield and proton peak position

The experimental thick target yield in the deuteron energy range of the observed yield plateau can be described as a sum of two components: the strongly energy dependent direct part and the constant contribution resulting from the thermal fusion. The corresponding formula can be presented as follows:

$$Y_{tot}(E) = Y_{scr}(E) + \mathcal{R}(T)_{exp} \tag{22}$$

Here, $Y_{scr}(E)$ describes the experimental dependence of the thick target yield determined at higher deuteron energies (see Eq. 13):

$$Y_{scr}(E) = M \exp\left(-\sqrt{\frac{E_G}{E+U_e}}\right) \tag{23}$$

which is fitted to the experimental data using the normalization factor $M$ and the screening energy $U_e$ as it is presented in Fig. 2a-4a and Fig. 7a. The thermal fusion contribution $\mathcal{R}(T)_{exp}$ can be then compared with theoretical calculations based on Eq. 21.

The peak position of 3 MeV protons could be determined with an uncertainty of 0.2 channel corresponding to about 3 keV by fitting a Gaussian function for measurements performed at higher deuteron energies, resulting in relatively high-count statistics. For lower energies, with fewer than 40 proton events, the mean value of the count distribution and its variance were calculated using the standard formulas:

$$\langle x\rangle = \sum_i n_i x_i \;/\; \sum_i n_i \qquad \text{and} \qquad \langle x^2\rangle = \sum_i n_i (x_i - \langle x\rangle)^2 \;/\; \sum_i n_i \tag{24}$$

where $x_i$ and $n_i$ are the channel number and the corresponding number of counts, respectively. The method is independent of the actual distribution function, which is an advantage since the line shape is not fully Gaussian. In this case, we set one-third of the variance as the uncertainty of the peak position being of about 0.5 channel, corresponding to 7.5 keV (see Fig. 2b-4b and 7b).

The experimentally determined proton peak position $x_p(E)$ could be then described by a weighted average of two values arising from reaction kinematics and a constant energy value of protons emitted from a resting CM system (thermal fusion):

$$x_p(E) = \frac{Y_{scr}(E)\cdot x_{kin}(E) + \mathcal{R}(T)_{exp}\cdot x_{kin}(0)}{Y_{scr}(E) + \mathcal{R}(T)_{exp}} \tag{25}$$

where $x_{kin}(E)$ can be obtained from the reaction kinematics:

$$x_{kin}(E) = a\cdot\left(\frac{3}{4}Q + \frac{1}{2}E(1+cos^2\varphi) + \sqrt{E}\,\cos\varphi\sqrt{\frac{3}{4}Q + \frac{1}{4}E(2+cos^2\varphi)}\right) + b \tag{26}$$

Here, $Q$ is the Q-value of the $^2H(d,p)^3H$ reaction, and φ is the detection angle with respect to the beam. The parameters $a$ and $b$ represent the linear energy calibration of the detector as slope and offset, respectively. The offset takes into account the proton energy loss in the aluminum protection foil placed in front of the Si-detector of about 22 keV corresponding to 1.5 channel. The detection angle φ could be independently fitted to experimentally determined proton peak positions (see the black "kinematic" curve in Fig. 2b and 7b). The obtained value of 138°±4° agrees very well with the actual angle.

### 5.4 Beam neutralization effect

Despite very good vacuum in the target chamber, the differential pumping system gradually reduces the vacuum pressure at the ion source of $10^{-6}$ mbar to the value of $5x10^{-9}$ mbar in the beam line behind the analyzing magnet. Due to increasing cross sections of charge exchange reactions on hydrogen molecules for lowering energies, a fraction of the ion beam can be neutralized in this part of the beam line and is not decelerated by the electric lenses in front of the target chamber. Considering 1.5 m distance between the electric beam steerer and the deceleration lenses and the neutralization cross section [28-30] of $10^{-15}$ cm$^2$ at $8x10^{-9}$ mbar rest gas pressure, the fraction of neutral beam can be estimated to about $6x10^{-5}$.

This value is very close to the experimentally determined fractions for the $D_2^+$ beam of approximately $10^{-4}$ (see Fig. 2a and 7a), which is especially important for measurements of strongly decreasing reaction cross sections.

Experimentally, we can observe the neutralization effect as a slightly increasing plateau in the reaction yield measured at extremely low energies since the neutralization cross section increases for lowering energies. The main argument for the neutralization origin of the observed $D_2^+$ yield plateau is the energy of emitted protons. The proton peak position is lower than would result from the decelerated beam energy; instead, it corresponds to the peak position of the high energy deuterons (Fig. 2b).

(a) (b)

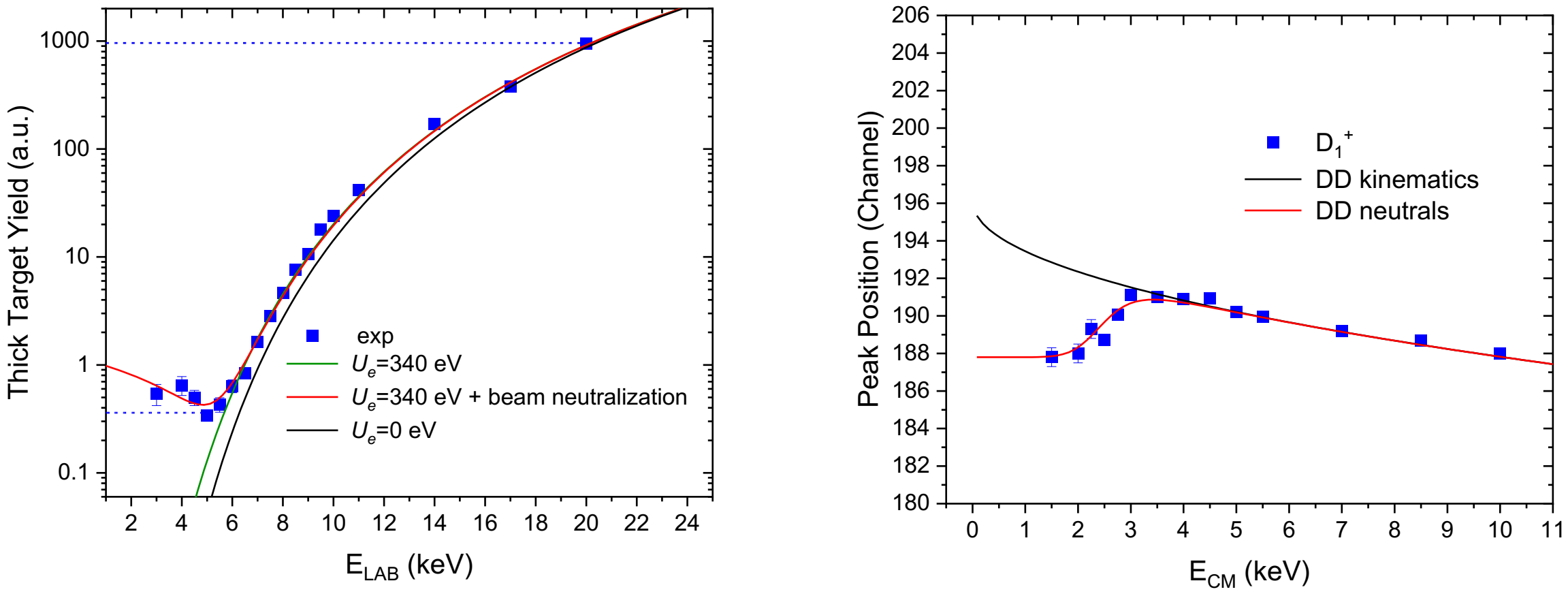

**Fig. 7: (a)** Thick target yield of the $^2H(d,p)^3H$ reaction measured on Zr using the atomic $D_1^+$ beam. The green full line corresponds to a theoretical curve obtained for the screening energy $U_e$=340$\pm$30 eV according to Eq. 19. The black full line represents the theoretical curve for $U_e$=0. The red curve results from slightly increasing neutralization cross section for lowering projectile energies and the green screening curve. The blue dashed lines show yield values obtained for the about $3x10^{-4}$ fraction of the neutralized $D_1^+$ beam (plateau) compared to the yield measured at 20 keV. **(b)** Channel number of the proton peak position. At the lowest deuteron energies, the experimental data obtained with the $D_1^+$ beam can be described by the neutral yield component corresponding to 20 keV deuterons. The related points are much lower than the theoretical values calculated according to the reaction kinematics (black curve). The red curve takes into account both the reaction kinematics and the neutralization effect.

Similar effects can also be observed for the atomic $D_1^+$ beam (see Fig. 7). A slightly increasing plateau can be observed towards lower deuteron energies with reaction yield values about two orders of magnitude higher (and higher beam energies) than in the case of the molecular $D_2^+$ beam, while the neutralization factor remains at the same level of about $3x10^{-4}$. In Fig. 7a, the “screened” yield functions corresponding to $U_e$=340eV (Eq. 19) and the curve without screening ($U_e$=0) are displayed, as well. The red full curve is a sum of $Y_{scr}(E)$ and the reaction yield resulting from the beam neutralization. The latter is assumed to be linearly decreasing with energy and was fitted to the data obtained for energies below 6 keV. The dashed blue horizontal lines in Fig. 7a serve to compare reaction yields at 20 keV and 6 keV and determine the neutralization factor. Slightly increasing plateau yield towards lowering projectile energies is in agreement with experimental data and theoretical calculations of the collisional recombination of $D_1^+$ and $D_2^+$ ions on deuterium $D_2$ molecules [29].

The beam neutralization process is supported by the observed proton peak position (see Fig. 7b). The proton energies measured for the beam decelerated below 6 keV are systematically lower than that expected from the reaction kinematics and reach the lowest value corresponding to that observed for 20 keV (CMS energy of 10 keV). The red curve in Fig. 7b is calculated as a weighted sum of two components: the "screened" yield function and the neutralization yield, similarly to the procedure applied to the thermal fusion yield (see Eq. 25).

## Data availability

All the experimental data are available from the corresponding authors upon request.

## Acknowledgements

The study is part of the CleanHME project from the European Union's Horizon 2020 research and innovation program under grant agreement No 951974. S.T. acknowledges support from the Anthropocene Institute.

## References

1. Spitaleri, C., Bertulani, C. A., Fortunato, L., Vitturi, A. The electron screening puzzle and nuclear clustering, *Phys. Lett. B* **755**, 275-278 (2016).
2. Czerski, K., Huke, A., Heide, P., Ruprecht, G. Enhancement of the electron screening effect for d+d fusion reactions in metallic environments, *Europhys. Lett.* **54**, 449-455, (2001).
3. Kasagi, J., Yuki, H., Baba, T., Noda, T., Taguchi, J., Shimokawa M., Galster W. Strongly Enhanced Li+D Reaction in Pd Observed in Deuteron Bombardment on $PdLi_x$ with Energies between 30 and 75 keV, *J. Phys. Soc. Jpn.* **73**, 608-612 (2004).
4. Cruz, J. et al. Electron screening in $^7Li(p,\alpha)\alpha$ and $^6Li(p,\alpha)^3He$ for different environments, *Phys. Lett. B* **624**, 181-185 (2005).
5. Cvetinovic, A. et al. Molecular screening in nuclear reactions, *Phys. Rev. C* **92**, 065801 (2015).
6. Ichimaru, S. Nuclear fusion in dense plasmas, *Rev. Mod. Phys.* **65**, 252-327 (1993).
7. Ichimaru, S. Pycnonuclear reactions in dense astrophysical and fusion plasmas, *Physics of Plasmas* **6**, 2649-2671 (1999).
8. Ouyed R., Fundamenski, W. R, Cripps, G. R, Sutherland, P. G. D–D Fusion in the Interior of Jupiter?, *Journal of Fusion Energy* **20**, 121–133 (2001).

9. Ouyed, R., Jaikumar, P. Nuclear fusion in the deuterated cores of inflated hot Jupiters, *Astrophys. Space Sci.* **361**, 89 (2016).
10. Czerski, K. et al. The $^{2}H(d,p)^{3}H$ reaction in metallic media at very low energies, *Europhys. Lett.* **68**, 363-369 (2004).
11. Huke, A. et al. Enhancement of deuteron-fusion reactions in metals and experimental implications, *Phys. Rev. C* **78**, 015803 (2008).
12. Kowalska, A. et al. Crystal Lattice Defects in Deuterated Zr in Presence of O and C Impurities Studied by PAS and XRD for Electron Screening Effect, *Materials* **16**, 6255 (2023).
13. Kowalska, A. et al. Electron Screening in Deuteron–Deuteron Reactions on a Zr Target with Oxygen and Carbon Contamination, *Materials* **18**, 1331 (2025).
14. Valat, M. et al. PAS characterization of defects produced by low-energy deuterons on Zr after carbon, oxygen and argon irradiations, to be published.
15. Czerski, K. et al. Screening and resonance enhancements of the 2H(d, p)3H reaction yield in metallic environments, *EPL* **113**, 22001 (2016).
16. Czerski, K. Deuteron-deuteron nuclear reactions at extremely low energies, *Phys. Rev. C* **106**, L011601 (2022).
17. Czerski, K. et al. Indications of electron emission from the deuteron-deuteron threshold resonance, *Phys. Rev. C* **109**, L021601 (2024).
18. Das Haridas, G. et al. High-energy electron measurements with thin Si detectors, *Measurement* **228**, 114392 (2024).
19. Dubey, R. et al. Experimental signatures of a new channel of the DD reaction at very-low energy, *Phys. Rev. X* **15**, 041004 (2025).
20. Sigmund, P. Energy density and time constant of heavy-ion-induced elastic-collision spikes in solids, *Appl. Phys. Lett.* **25**, 169–171 (1974).
21. Vineyard, G. H. Thermal spikes and activated processes, *Radiation Effects* **29**, 245–248 (1976).
22. Sigmund, P., Claussen, C. Sputtering from elastic-collision spikes in heavy-ion-bombarded metals, *J. Appl. Phys.* **52**, 990–993 (1981).
23. Dufour, C., Toulemonde, M. *Models for the Description of Track Formation* in W. Wesch and E. Wendler (eds.), *Ion Beam Modification of Solids*, Springer Series in Surface Sciences **61**, 63–104 (2016).
24. Miotello, A. Kelly, R. Revisiting the thermal-spike concept in ion-surface interactions, *Nucl. Instrum. Methods Phys. Res. B* **122**, 458–469 (1997).
25. Lin, Z., Johnson, R. A., Zhigilei, L. V. Computational study of the generation of crystal defects in a bcc metal target irradiated by short laser pulses, *Phys. Rev. B* **77**, 214108 (2008).
26. He, M. et al. Computational Study of Short Pulse Laser Induced Generation of Crystal Defects in Ni-Based Single-Phase Binary Solid Solution Alloys, *J. Phys. Chem. C* **123**, 09922 (2019).
27. Arai, K. et al. Tensor Force Manifestations in Ab Initio Study of the $^{2}H(d,\gamma)^{4}He$, $^{2}H(d,p)^{3}H$, and $^{2}H(d,n)^{3}He$ Reactions, *Phys. Rev. Lett.* **107**, 132502 (2011).
28. Phelps, A. V. Cross Sections and Swarm Coefficients for $H^{+}$, $H_2^{+}$, $H_3^{+}$, H, $H_2$, and $H^{-}$ in $H_2$ for Energies from 0.1 eV to 10 keV, *Journal of Physical and Chemical Reference Data* **19**, 653-675 (1990).
29. McClure, G. W. Charge Exchange and Dissociation of $H^{+}$, $H_2^{+}$ and $H_3^{+}$ Ions Incident on $H_2$ Gas, *Phys. Rev.* **130**, 1852-1859 (1963).
30. Tabata, T., Shiray, T. Analytic Cross Sections for Collisions of H+, H+2, H+3, H, H2, and H− with Hydrogen Molecules, *Atomic Data and Nuclear Data Tables* **76**, 1–25 (2000).

31. Czerski, K. et al. 2024, arXiv:2409.02112 [nucl-ex].
32. Ziegler, J. F., Biersack, J. P., Littmark, U. *The Stopping and Range of Ions in Solids*, vol. 1 of *The Stopping and Ranges of Ions in Matter*, Pergamon Press, New York (1985).
33. Iliadis, C. *Nuclear Physics in Stars*, Willey-VCH Verlag GmbH & Co. KGaA, Weinheim (2007).
34. Tsuchiya, B. et al. Isotope Effects in Thermal Diffusivity and Electrical Resistivity of Zirconium Hydride and Deuteride, *Fusion Science and Technology* **39,** 402-406 (2002).
35. Tsuchiya, B. et al. Heat conductions due to electrons and phonons for titanium hydride and deuteride, *Journal of Alloys and Compounds* **356–357**, 223–226 (2003).
36. Zhen Tong et al. Comprehensive first-principles analysis of phonon thermal conductivity and electron-phonon coupling in different metals, *Phys. Rev. B* **100**, 144306 (2019).
37. Olsson, P. A. T. et al., Ab initio thermodynamics of zirconium hydrides and deuterides, *Computational Materials Science* **86**, 211–222 (2014).
38. Hu, C. H. et al. First-principles investigations of isotope effects in thermodynamic properties of TiX2 (X =H, D, and T) system, *Journal of Alloys and Compounds* **450**, 369–374 (2008).
39. Ali, M. L., Ramos, S. B., Fernandez Guillermet, A. Heat capacity and enthalpy of palladium: A critical analysis of experimental information, *Calphad* **84**, 102670 (2024).
40. Long, X. et al., Hydrogen Isotope Effects of Ti, Zr Metals, *Fusion Science and Technology* **60,** 1568-1571 (2011).
41. Kaess, U. et al. Hydrogen and deuterium diffusion in titanium dihydrides/dideuterides", *Journal of Alloys and Compounds* **259**, 74-82 (1997).
42. Zhou, X. W. et al. Temperature- and concentration-dependent hydrogen diffusivity in palladium from statistically-averaged molecular dynamics simulations, *Scripta Materialia* **149**, 103–107 (2018).
43. Kasagi, J. et al. Strongly Enhanced DD Fusion Reaction in Metals Observed for keV D+ Bombardment, *J. Phys. Soc. Jpn.* **71**, 2881-2885 (2002).
44. Raiola, F. et al., Electron screening in d(d,p)t for deuterated metals and the periodic table, *Phys. Lett. B* **547**, 193-199 (2002).
45. Schiwietz, G. et al. Evidence for an ultrafast breakdown of the BeO band structure due to swift argon and xenon ions, *Phys. Rev. Lett. 105, 187603 (2010).*
46. Lipson, A.G., Rusetskii, A.S., Karabut, A.B., Miley, G. DD Reaction Enhancement and X-Ray Generation in a High-Current Pulsed Glow Discharge in Deuterium with Titanium Cathode at 0.8-2.45 kV, *Journal of Experimental and Theoretical Physics* **100**, 1334-1349 (2005).
47. Schenkel, T. et al. Investigation of light ion fusion reactions with plasma discharges, *J. Appl. Phys.* **126**, 203302 (2019).
48. Duff, W. H., Zhigilei, L. V. Computational study of cooling rates and recrystallization kinetics in short pulse laser quenching of metal targets, *J. Phys.: Conf. Ser.* **59**, 413–417 (2007).
49. Rymzhanov, R. A. et al. Recrystallization as the governing mechanism of ion track formation, *Sci. Rep.* **9**, 3837 (2019).
50. Sun, G., Xu, J., Harrowell, P. The mechanism of the ultrafast crystal growth of pure metals from their melts, *Nat. Mater.* **17**, 881–886 (2018).
51. Karahadian, M. E., Colborne, M., Persaud, A., Schenkel, T., Munday, J. N. Enhanced nuclear fusion in the sub-keV energy regime, *Nature Comm.,* in press (2026).
52. Kaczmarski, M., et al. New accelerator facility for measurements of nuclear reactions at energies below 1 keV, *Acta Phys. Pol. B* **45,** 509-518 (2014).